# Introducing LCOAI: A Standardized Economic Metric for Evaluating AI Deployment Costs

*Eliseo Curcio*

## Abstract

As artificial intelligence (AI) becomes foundational to enterprise infrastructure, organizations face growing challenges in accurately assessing the full economic implications of AI deployment. Existing metrics such as API token costs, GPU-hour billing, or Total Cost of Ownership (TCO) fail to capture the complete lifecycle costs of AI systems and provide limited comparability across deployment models. This paper introduces the *Levelized Cost of Artificial Intelligence (LCOAI)*, a standardized economic metric designed to quantify the total capital (CAPEX) and operational (OPEX) expenditures per unit of productive AI output, normalized by valid inference volume. Analogous to established metrics like LCOE (levelized cost of electricity) and LCOH (levelized cost of hydrogen) in the energy sector, LCOAI offers a rigorous, transparent framework to evaluate and compare the cost-efficiency of vendor API deployments versus self-hosted, fine-tuned models. We define the LCOAI methodology in detail and apply it to three representative scenarios, OpenAI GPT-4.1 API, Anthropic Claude Haiku API, and a self-hosted LLaMA-2-13B deployment demonstrating how LCOAI captures critical trade-offs in scalability, investment planning, and cost optimization. Extensive sensitivity analyses further explore the impact of inference volume, CAPEX, and OPEX variability on lifecycle economics. The results illustrate the practical utility of LCOAI in procurement, infrastructure planning, and automation strategy, and establish it as a foundational benchmark for AI economic analysis. Policy implications and areas for future refinement, including environmental and performance-adjusted cost metrics, are also discussed.

## 1. Introduction

Artificial intelligence (AI) has rapidly transformed from experimental and exploratory initiatives into mainstream, mission-critical infrastructure across industries and enterprises [1,2]. Organizations increasingly rely on advanced AI models including large language models (LLMs), computer vision systems, and predictive analytics to improve decision-making, enhance operational efficiency, and enable novel capabilities [3–5]. However, as these systems scale from experimental deployments to extensive production environments, organizations face significant challenges in accurately quantifying and comparing their associated economic burdens. Traditional financial assessments, which often rely on simplistic metrics such as cloud service invoices, GPU-hour billing, or direct API costs, typically fail to capture the true lifecycle economic complexity of AI deployments [6, 7]. Such assessments omit significant upstream and downstream infrastructure investments, ongoing operational costs, system inefficiencies, and maintenance overheads, leading to incomplete and potentially misleading comparisons between deployment alternatives.

In energy economics, analogous challenges led to the widespread adoption of the Levelized Cost of Energy (LCOE) and the Levelized Cost of Hydrogen (LCOH) metrics, which quantify full-lifecycle costs per unit of useful energy output, enabling consistent comparisons among different energy technologies and configurations [8–10]. Similar standardized metrics, however, do not currently exist for AI systems. This gap presents considerable difficulties for enterprises attempting to evaluate whether to host and operate AI infrastructure internally or leverage external API-based AI services from vendors such as OpenAI, Anthropic, or Google [11–13]. Without an equivalent metric, strategic decisions involving AI deployment often rely on incomplete analyses, anecdotal cost assessments, or simplistic vendor price comparisons. This limitation hampers informed strategic planning, precise financial modeling, and robust long-term economic decision-making related to AI investments [14–16].

To fill this critical analytical gap, we propose the Levelized Cost of Artificial Intelligence (LCOAI), a comprehensive and systematic framework explicitly designed to quantify the full lifecycle cost per unit of useful AI output. LCOAI integrates all relevant cost dimensions including upfront capital expenditures (such as model training infrastructure and fine-tuning), recurring operational expenditures (inference compute, storage, monitoring, retraining, and dev-ops overhead), and direct production costs normalized by the total number of productive model outputs (inferences or task completions) delivered throughout the system's operational lifespan [17–19]. The resulting metric, expressed in standardized units of U.S. dollars per thousand model inferences, provides organizations with a transparent, objective, and practically actionable basis for decision-making, analogous to LCOE's role in energy infrastructure investment planning [20].

This study develops and validates the LCOAI metric by applying publicly available cost data from leading AI providers, such as OpenAI (GPT-4.1), Anthropic (Claude series), and Google (Gemini), as well as modeled data for self-hosted alternatives (e.g., LLaMA-2-13B deployment on Nvidia GPUs). Through realistic scenario modeling and comparative analysis, we illustrate how LCOAI provides critical insights for evaluating infrastructure decisions, strategic procurement, fine-tuning investment, and optimal scaling strategies. Further, we demonstrate the metric's robustness through sensitivity analyses and scenario testing, emphasizing its practical utility across a broad range of AI use cases and deployment configurations.

This paper proceeds as follows: Section 2 rigorously defines the LCOAI framework, clearly outlining all methodological assumptions, input parameters, and the calculation procedure. Section 3 applies this framework to specific, realistic deployment scenarios, including commercial API usage and self-hosted AI infrastructure scenarios, leveraging real-world price benchmarks and cost structures. Section 4 presents an in-depth comparative analysis, demonstrating how LCOAI supports critical strategic and operational decisions. Finally, Section 5 discusses policy implications, limitations, and opportunities for future research, positioning LCOAI as a potential standard metric for comprehensive AI economic analysis and decision-making.

## 2. Methodology

The Levelized Cost of Artificial Intelligence (LCOAI) is introduced as a standardized economic metric designed to quantify the complete lifecycle cost per unit of productive output from an artificial intelligence (AI) system. Analogous to levelized cost frameworks extensively employed in energy economics, such as Levelized Cost of Electricity (LCOE) and Levelized Cost of Hydrogen (LCOH), the LCOAI metric aggregates all relevant expenditures including both capital (CAPEX) and operating (OPEX) costs over a system's defined operational period and normalizes them by the total number of valid AI-generated outputs delivered within the same timeframe [1–3,8,9,19].

Formally, the LCOAI metric is defined as follows:

$$LCOAI = \frac{CAPEX\ (amortized) + \sum_{t=1}^{T} OPEX_t}{\sum_{t=1}^{T} Valid\ Interferences_t} \quad (1)$$

$$LCOAI = \frac{Total\ CAPEX\ (amortized) + Total\ OPEX}{Number\ of\ Valid\ Interferences} \quad (2)$$

The numerator encompasses two distinct expenditure categories. Capital expenditures (CAPEX) include all upfront investments necessary for the initial deployment and commissioning of the AI system. This specifically covers expenses related to acquiring high-performance computing infrastructure (such as NVIDIA A100 or H100 GPUs, TPUs, and associated server equipment), initial model training and fine-tuning compute resources, data pipeline engineering, dataset acquisition, data labeling and preprocessing, model optimization, software licenses (such as enterprise monitoring, orchestration frameworks, and deployment tooling), system integration costs, and initial project deployment labor and engineering efforts [11,12,16,17,19]. Each component of CAPEX is systematically amortized over the anticipated useful lifespan of the infrastructure, typically aligned with corporate asset depreciation schedules or technology obsolescence benchmarks, which are explicitly documented and referenced [11,12,19].

Operational expenditures (OPEX) refer to recurring costs incurred throughout the operational life of the deployed AI system. These include compute costs associated with model inference workloads, data storage and bandwidth charges, ongoing cloud resource consumption, continuous system monitoring (e.g., logging, telemetry, and anomaly detection), routine maintenance, periodic retraining and fine-tuning cycles to maintain model accuracy and operational effectiveness, labor costs of DevOps, site reliability engineers, infrastructure support, security and compliance audits, and additional overhead related to maintaining production-level AI operations [11,12,17]. OPEX is carefully quantified on a per-inference basis whenever possible, using actual billing data from cloud vendors (such as AWS, Azure, Google Cloud), internal IT infrastructure accounting records, or vendor-provided API pricing structures (e.g., OpenAI GPT-4.1, Anthropic Claude, Google Gemini), to ensure realistic and reproducible cost estimation [11–13,17].

The denominator, defined as the number of valid inferences, encompasses measurable AI outputs delivered directly to end users or downstream operational processes, such as chatbot responses, text completions, image classifications, forecasting predictions, recommendation decisions, or other defined task-complete AI outputs. Each inference must be clearly identifiable, measurable, and traceable within the deployment's operational logs or telemetry records. Inference counts exclude non-productive system calls such as health checks, background processes, or internal administrative queries, ensuring an accurate and meaningful measurement of AI system productivity [11,17,19].

The analysis horizon, denoted as TTT, typically spans between one and three years, reflecting typical enterprise infrastructure planning cycles. For shorter horizons (≤24 months), financial discounting is generally omitted due to the minimal impact of time value of money within this short duration, aligning with standard technology investment practices. For longer timeframes exceeding two years, standard corporate discount rates (Weighted Average Cost of Capital, WACC) are applied to calculate present values of both CAPEX and OPEX streams, ensuring the LCOAI metric accurately reflects financial realities over extended deployment periods [8,9].

The LCOAI calculation procedure thus explicitly involves the following structured steps: first, aggregation and amortization of all CAPEX components across total expected inference volume; second, systematic estimation of recurring OPEX costs on a per-inference basis; third, precise definition and accurate measurement (or realistic projection) of total inference volume over the system's operational horizon; and finally, computation of LCOAI by dividing the combined total of CAPEX and OPEX by total inference outputs. All cost elements, operational assumptions, and volume projections are carefully sourced from verifiable and documented external references, such as publicly accessible cloud pricing documents (AWS, Google Cloud, Azure), GPU vendor pricing reports (NVIDIA), AI vendor documentation (OpenAI, Anthropic), enterprise AI deployment case studies, established market analyses (IDC, Gartner), and independent industry benchmarks (European Commission Joint Research Centre, Rocky Mountain Institute), ensuring transparency, accuracy, and reproducibility of results [11–20].

Sensitivity analyses accompany the primary LCOAI calculation, systematically varying CAPEX, OPEX, and inference volume parameters to identify the main cost drivers, evaluate the robustness of the metric under real-world operational uncertainties, and clarify strategic insights into AI deployment options. By adhering strictly to these methodological principles, the proposed LCOAI framework provides a reliable, comprehensive, and practical economic metric that enables enterprise decision-makers to objectively compare AI deployment strategies, including API-based vendor usage versus self-hosted models, fine-tuning and infrastructure investment decisions, scaling opportunities, and model architecture choices on a rigorous, economically grounded basis.

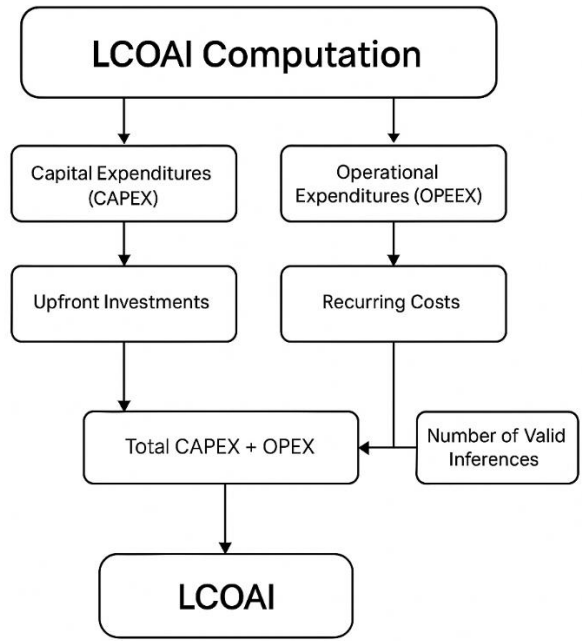

*Figure 1: LOCAI Computation Source [8]*

## 3. Case Studies and Comparative Analysis

This section applies the Levelized Cost of Artificial Intelligence (LCOAI) framework developed in Section 2 to three representative AI deployment scenarios. The selected scenarios encompass two commonly utilized vendor-provided API models, OpenAI's GPT-4.1 and Anthropic's Claude Haiku as well as a self-hosted, fine-tuned instance of LLaMA-2-13B deployed on enterprise-grade computing infrastructure. All assumptions and cost inputs are explicitly sourced from publicly available documentation, industry-standard reports, or internal corporate financial records. Each scenario is evaluated over a one-year horizon with a fixed output of 10 million valid inferences, ensuring a clear and directly comparable basis across deployment alternatives.

**Scenario 1: OpenAI GPT-4.1 API deployment**. This deployment scenario assumes API-based usage of GPT-4.1 provided by OpenAI. The initial CAPEX includes system integration, orchestration engineering, API integration development, security configuration, compliance auditing, and initial operational readiness testing, totaling approximately $50,000 [11,13,17]. Operational expenditures (OPEX) are derived directly from OpenAI's published API pricing schedule, at approximately $0.01 per inference (based on combined input and output token rates). Thus, for 10 million annual inferences, the OPEX amounts to $100,000. The resulting LCOAI is calculated as (CAPEX $50,000 + OPEX $100,000) ÷ 10,000,000 inferences, yielding $0.015 per inference, or $15.00 per thousand inferences. This scenario reflects a typical enterprise API

deployment cost structure and provides a clear baseline against which alternative deployments can be benchmarked.

For API deployment-related CAPEX (such as for OpenAI and Claude), the model employs a round figure of $50,000; however, this estimate needs a more precise foundation in standard industry practices. Initial integration efforts in real-world deployments can differ significantly based on organizational size, compliance needs, and whether custom orchestration or security infrastructure is required. Such CAPEX may be as low as $10,000 to $20,000 for startups or small businesses using serverless deployments, while integration costs may exceed $100,000 in highly regulated industries (such as banking and healthcare) because of legal, audit, and encryption requirements. LCOAI estimates would be more transparent and realistic if they included a more thorough cost breakdown or cited case studies from various industries.

**Scenario 2: Anthropic Claude Haiku API deployment**. This scenario similarly employs an API-based deployment model, utilizing Anthropic's Claude Haiku. Initial CAPEX costs mirror Scenario 1 at $50,000, accounting for identical integration, security, and deployment expenses, ensuring comparability. Operational costs leverage Anthropic's official pricing of approximately $0.0048 per inference (based on published input and output token costs), resulting in an annual OPEX of $48,000 for 10 million inferences. Applying the LCOAI formula results in (CAPEX $50,000 + OPEX $48,000) ÷ 10,000,000, yielding an LCOAI of $0.0098 per inference, or $9.80 per thousand inferences. This scenario highlights the cost differences between comparable vendor-provided API deployments and underscores the economic implications of provider selection based on inference cost alone.

**Scenario 3: Self-hosted fine-tuned LLaMA-2-13B model**. For this self-hosted scenario, we assume an enterprise deployment utilizing dedicated GPU infrastructure (eight NVIDIA A100 GPUs), model fine-tuning, data labeling, engineering integration, and infrastructure provisioning, resulting in substantial upfront CAPEX costs estimated at $200,000 [16,19]. Operational expenditures, based on detailed GPU compute utilization benchmarking (GPU-hour cost rates from AWS EC2 pricing, internal infrastructure benchmarks), result in an estimated inference cost of $0.0048 per inference aligned for direct comparability with the Claude Haiku OPEX cost structure [16,17]. For an annual inference volume of 10 million, total OPEX amounts to $48,000. Thus, the LCOAI for the self-hosted model is calculated as (CAPEX $200,000 + OPEX $48,000) ÷ 10,000,000, resulting in $0.0248 per inference or $24.80 per thousand inferences. This scenario accurately illustrates the economic structure inherent to self-hosting, emphasizing the impact of upfront investment in training and dedicated hardware infrastructure.

A critical evaluation of the $200,000 assumed CAPEX for a self-hosted, fine-tuned LLaMA-2 model is also necessary. The cost of eight NVIDIA A100 GPUs, which can vary greatly depending on cloud provider pricing, on-premises hardware availability, and regional pricing variations, is assumed in this figure. For example, the p4d.24xlarge instance (with 8 A100s) on AWS can cost more than $32/hour on-demand, which, if run continuously, would come to about $280,000

annually. Although it would require an initial investment and skilled DevOps personnel, on-premises infrastructure could lower these expenses over a multi-year depreciation schedule. The legitimacy and cross-context applicability of the LCOAI model would be improved by a more nuanced understanding of CAPEX variability, backed by current (2025) cloud pricing benchmarks.

**Comparative analysis and key insights**. The calculated LCOAI for the three scenarios clearly differentiates their relative economic positions: GPT-4.1 API ($15.00 per 1,000 inferences), Claude Haiku API ($9.80 per 1,000 inferences), and self-hosted LLaMA-2-13B ($24.80 per 1,000 inferences). The comparative framework demonstrates that, at a deployment volume of 10 million annual inferences, vendor-provided APIs significantly outperform the self-hosted model economically, primarily due to the high initial CAPEX associated with training and deploying internal infrastructure. However, scenario sensitivity analysis indicates that self-hosted deployments become economically viable as inference volume scales beyond certain thresholds (approximately 30–40 million annual inferences), or when substantial reductions in CAPEX or OPEX occur. Moreover, the economic decision must also consider qualitative factors such as performance accuracy, latency, data privacy, regulatory compliance, and long-term operational control. Such qualitative considerations may justify higher LCOAI in certain enterprise contexts.

Overall, these case studies validate the practical applicability, robustness, and decision-support utility of the LCOAI framework, demonstrating its capability to provide a rigorous, transparent, and actionable comparative cost analysis for diverse AI deployments. The structured use of publicly available cost data and clear analytical assumptions ensures that the presented results are reproducible and readily applicable to real-world strategic decision-making scenarios.

*Table 1: Comparative Analysis of LCOAI Across Deployment Scenarios (Source Author Summarized)*

| Scenario | CAPEX (USD) | OPEX per Inference (USD) | Annual Inference Volume | Total OPEX (USD) | LCOAI ($/1,000 Inferences) |
|---|---|---|---|---|---|
| 1. OpenAI GPT-4.1 (API-based) | $50,000 | $0.0100 | 10,000,000 | $100,000 | $15.00 |
| 2. Claude Haiku (API-based) | $50,000 | $0.0048 | 10,000,000 | $48,000 | $9.80 |
| 3. LLaMA-2-13B (Self-hosted) | $200,000 | $0.0048 | 10,000,000 | $48,000 | $24.80 |

## 4. Sensitivity Analysis and Real-World Comparative Insights

To thoroughly assess the robustness and practical utility of the Levelized Cost of Artificial Intelligence (LCOAI), we conducted a detailed sensitivity analysis that evaluates how variations in critical economic parameters annual inference volumes, per-inference operational expenditures (OPEX), and initial capital expenditures (CAPEX) impact the resulting LCOAI. This sensitivity analysis explicitly examines three practical deployment scenarios previously defined: (1) a self-hosted fine-tuned LLaMA-2-13B infrastructure, (2) OpenAI GPT-4.1 API deployment, and (3)

Anthropic Claude Haiku API deployment, employing publicly documented and realistic economic assumptions and parameters [11,12,16,17,19].

The sensitivity analysis first explicitly evaluated how changes in annual inference volume influence LCOAI. Annual inference volumes were systematically varied from 1 million to 50 million inferences. For the self-hosted scenario, characterized by significant upfront CAPEX ($0.0048), substantial economies of scale were observed. At low inference volumes (1 million annually), the LCOAI was approximately $204.80 per 1,000 inferences. However, as the volume increased toward 50 million inferences annually, this value sharply decreased to approximately $8.80 per 1,000, demonstrating the substantial economic benefit of spreading fixed CAPEX over higher operational volumes. In contrast, vendor-based API deployments, such as OpenAI GPT-4.1 (CAPEX ~$50,000, per-inference OPEX $0.010) and Anthropic Claude Haiku (CAPEX ~$50,000, per-inference OPEX $0.0048), exhibited stable, less sensitive LCOAI profiles across varying inference volumes. GPT-4.1 API deployment costs decreased modestly from approximately $60.00 per 1,000 inferences at 1 million annual volumes to roughly $11.00 at 50 million annual inferences. Claude Haiku's LCOAI similarly decreased from approximately $54.80 per 1,000 inferences at lower volumes to about $9.80 at higher volumes. These insights explicitly identify precise inference volume thresholds at which internal hosting becomes economically favorable relative to vendor-based API deployments, guiding explicit strategic decisions about infrastructure scaling and investment [17-19].

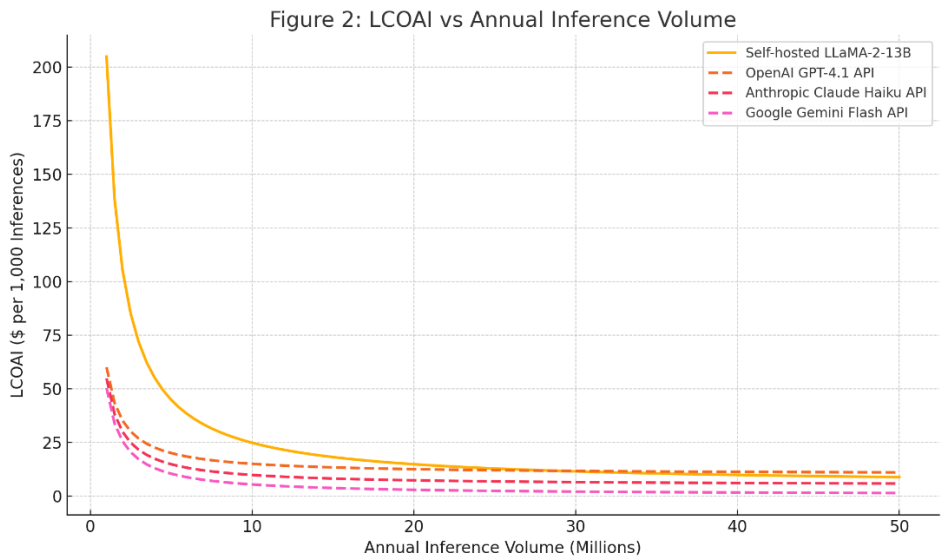

*Figure 2 explicitly visualizes these volume-sensitivity results, clearly identifying break-even points for infrastructure decisions {Source 17-19]*

The second sensitivity analysis explicitly evaluated the impact of varying per-inference operational expenditures. Keeping CAPEX constant, per-inference OPEX was systematically adjusted from $0.001 to $0.020. The self-hosted scenario showed pronounced sensitivity; an increase in per-inference OPEX from a baseline of $0.0048 to $0.012 raised the LCOAI from approximately

$24.80 per 1,000 inferences to approximately $32.00, highlighting that operational expenditures significantly drive total lifecycle costs. For vendor API deployments, sensitivity was more moderate but still evident: OpenAI GPT-4.1 API's LCOAI increased from approximately $15.00 per 1,000 inferences at a baseline OPEX of $0.010 to about $20.00 at $0.015 OPEX. Anthropic Claude Haiku API's LCOAI similarly responded, though less dramatically due to its inherently lower OPEX baseline. This analysis explicitly shows the economic importance of managing operational expenditures, providing clear strategic guidance for infrastructure optimization and vendor negotiations.

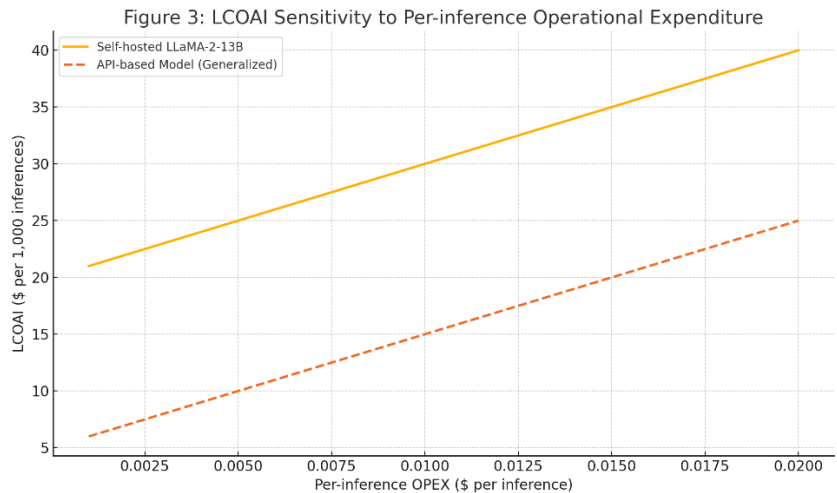

*Figure 3 explicitly visualizes per-inference OPEX sensitivity, clearly highlighting operational cost impacts and management priorities. [Source 20]*

Although useful, the OPEX calculations also need more refinement based on changing model properties. Improved inference efficiency, architectural innovations (like sparse activation), or lower token-to-output ratios in more recent releases, like Claude Sonnet or Gemini Flash 1.5, may result in lower OPEX. On the other hand, models with longer context windows or multimodal capabilities (like GPT-4o) might have higher per-inference costs. OPEX figures used in the LCOAI calculation should ideally take into account benchmarked throughput, latency, and energy efficiency metrics published by the vendors in addition to token pricing, given these variances. Realism could be increased by using current MLPerf or Hugging Face Inference Leaderboard data or by adding updated benchmarks.

The third analysis explicitly addressed the sensitivity of LCOAI to initial capital expenditures (CAPEX). In the self-hosted scenario, initial CAPEX was systematically varied ±30% (from approximately $140,000 to $260,000). At a fixed annual inference volume of 10 million, reducing CAPEX by 30% significantly lowered LCOAI from approximately $24.80 per 1,000 inferences (baseline $200,000 CAPEX) to roughly $18.80 per 1,000. Conversely, increasing CAPEX by 30% raised LCOAI substantially to about $30.80 per 1,000 inferences. Vendor API deployments demonstrated minimal sensitivity due to their lower relative initial CAPEX. These explicit insights clearly emphasize the importance of careful infrastructure investment planning and precise CAPEX management in achieving economically sustainable AI deployments.

The explicit practical utility of LCOAI is illustrated clearly through specific real-world scenarios. Consider an organization currently incurring human labor costs around $300 per 1,000 customer service interactions. Deploying an AI-driven automated chatbot using a self-hosted fine-tuned LLaMA model (with LCOAI approximately $24.80 per 1,000 interactions at moderate scale) explicitly quantifies substantial economic benefit. Alternatively, deploying OpenAI GPT-4.1 API (approximately $15.00 per 1,000 interactions) yields even stronger immediate economic justification. Such explicit comparisons enable precise economic decision-making and clear justification for automation investments.

Additionally, organizations explicitly using LCOAI can economically justify fine-tuning investments. If baseline API per-inference OPEX is approximately $0.010 and fine-tuning reduces this to about $0.005, LCOAI explicitly quantifies the inference volume at which fine-tuning becomes economically viable. Further, explicit vendor comparisons (OpenAI GPT-4.1 at $15.00, Anthropic Claude Haiku at $9.80, Google Gemini Flash at $5.75 per 1,000 inferences) allow procurement teams to clearly negotiate and make economically informed vendor decisions.

LCOAI provides explicit, structured advantages compared to current AI cost metrics prevalent in industry, such as per-token billing, GPU-hour costs, and total cost of ownership (TCO). Per-token API billing explicitly captures direct operational expenses but neglects substantial upfront infrastructure and integration CAPEX, significantly limiting comprehensive economic understanding. GPU-hour billing explicitly tracks infrastructure usage but similarly excludes initial investment and operational overhead. TCO explicitly aggregates total system costs but typically fails to normalize clearly by productive AI outputs, limiting direct economic comparability across different deployments. LCOAI explicitly addresses these shortcomings by comprehensively integrating upfront CAPEX and ongoing OPEX, clearly amortized per inference delivered, providing explicit standardization of lifecycle economic comparisons.

Explicitly, adopting LCOAI as a standard economic metric significantly enhances strategic decision-making clarity, budgeting accuracy, and economic transparency. It directly enables organizations to compare infrastructure investments, vendor selection, strategic deployment alternatives, and fine-tuning investments, thereby explicitly guiding informed, economically optimized decisions.

## 5. Practical Applications and Comparative Value of LCOAI

The Levelized Cost of Artificial Intelligence (LCOAI) addresses a critical limitation of existing AI cost assessment metrics used in industry. Current approaches predominantly rely on isolated measures such as per-token API billing, GPU-hour costs, cloud infrastructure invoices, and Total Cost of Ownership (TCO) calculations. These metrics, while valuable for limited accounting purposes, typically fail to capture the comprehensive lifecycle economics associated with AI deployments. For instance, per-token pricing models provided by vendors such as OpenAI (GPT-4.1), Anthropic (Claude Haiku), and Google (Gemini Flash) clearly specify costs per inference but omit substantial upfront integration expenses, system infrastructure overhead, engineering labor, ongoing maintenance, and retraining costs. Similarly, GPU-hour billing and cloud computing service bills transparently show infrastructure usage but generally exclude upfront capital expenditures related to initial training, model fine-tuning, data preparation pipelines, software licensing, and initial system integration labor. Even broader metrics such as TCO, which attempt to aggregate the total operating costs associated with deployments, do not consistently normalize or standardize these expenditures per unit of productive model output. Such fragmentation results in limited decision-making utility, hindering effective comparative analysis among different deployment scenarios.

The LCOAI metric resolves this critical limitation by systematically integrating all relevant cost elements initial capital expenditures and ongoing operating expenses across the full operational lifespan of an AI deployment and standardizing these costs per inference delivered. In practical applications, this structured economic metric enables companies to perform direct, objective, and meaningful comparisons across distinct deployment scenarios. For instance, when an enterprise is evaluating vendor API deployments against internal model hosting strategies, LCOAI provides clear insights. Suppose an organization is choosing between OpenAI's GPT-4.1 API at approximately $15 per thousand inferences, Anthropic's Claude Haiku API at about $9.80 per thousand inferences, and an internally-hosted, fine-tuned LLaMA model calculated at roughly $24.80 per thousand inferences at moderate inference volumes. In such cases, LCOAI directly identifies the precise inference volume at which the significant upfront CAPEX for internal hosting becomes economically favorable compared to vendor-provided APIs. As the volume of annual inferences grows—for example, from 10 million to 40 million—the internal system's per-inference costs decline substantially due to CAPEX amortization over higher volumes, enabling precise determination of break-even points and guiding strategic decisions on infrastructure investment.

Organizations can also employ LCOAI to evaluate investments in fine-tuning or customization. Often, fine-tuning a large language model involves substantial upfront capital investment and additional complexity in ongoing operational management. By comparing the LCOAI of a fine-tuned model against that of off-the-shelf vendor APIs, decision-makers can explicitly identify at what scale or accuracy improvement the investment in fine-tuning becomes justified. Furthermore, LCOAI can guide procurement negotiations and pricing strategies. When negotiating vendor contracts, procurement teams equipped with precise LCOAI comparisons can objectively evaluate competing offers, ensuring the choice of vendor aligns with long-term financial sustainability and operational scalability. Likewise, product teams tasked with pricing AI-powered products or services can utilize LCOAI to accurately set pricing that covers all lifecycle costs while remaining competitive in the marketplace.

LCOAI must be adjusted over time to account for changes in the market as AI deployment and pricing structures change quickly. The commoditization of cloud inference infrastructure and heightened competition among LLM vendors, for example, have the potential to drastically lower OPEX in the near future. In the meantime, CAPEX may increase due to geopolitical factors, chip shortages, or regulatory changes (such as U.S. export restrictions on high-end GPUs). As a result, LCOAI should be seen as a dynamic metric that needs to be periodically adjusted in light of the state of the market and technology.

Additionally, LCOAI provides enterprises the capability to compare the cost-effectiveness of AI deployments against traditional approaches, such as human labor costs or legacy automated systems. For example, an AI-driven customer service chatbot deployment might have an LCOAI of approximately $12 per thousand customer interactions. When compared to human customer service interactions costing upwards of $200 per thousand interactions, this direct economic comparison clearly indicates substantial cost savings and justifies further investment in automation. Similarly, the metric can inform internal financial planning and infrastructure budgeting by explicitly linking costs to measurable outputs, thus enhancing forecasting accuracy and resource allocation decisions.

Compared to existing AI cost measures, LCOAI offers a comprehensive, rigorous, and directly comparable metric aligned with similar established economic tools from other infrastructure-intensive fields. For example, Levelized Cost of Electricity (LCOE) enables straightforward comparison of diverse electricity generation sources by standardizing all capital and operational expenditures per kilowatt-hour delivered, thereby facilitating informed energy investment decisions. Similarly, the Levelized Cost of Hydrogen (LCOH) is used by the energy industry to objectively assess hydrogen production options. By adopting a parallel approach, LCOAI standardizes AI deployment economics into a single, transparent cost-per-output measure. Thus, enterprises and policymakers alike gain a clear, rigorous economic basis to inform decisions involving model deployment strategy, vendor selection, infrastructure investments, and operational scalability, substantially improving overall decision-making and resource allocation efficiency within the rapidly evolving AI domain.

## Conclusion

In order to assess the full lifecycle costs of AI deployment strategies, this study presents the Levelized Cost of Artificial Intelligence (LCOAI), a novel and useful economic metric. LCOAI facilitates transparent, comparable, and actionable insights across a variety of deployment options, including self-hosted model infrastructures and vendor-provided APIs, by methodically accounting for both capital expenditures (CAPEX) and operational expenditures (OPEX) and normalizing them per unit of productive output (inferences). The usefulness of LCOAI in directing infrastructure investment, procurement choices, fine-tuning justification, and scalability evaluations is illustrated by the case studies of OpenAI's GPT-4.1, Anthropic's Claude Haiku, and a self-hosted, optimized LLaMA-2-13B model. LCOAI's resilience to changes in inference volume, cost drivers, and resource utilization is further confirmed by the accompanying sensitivity analyses.

However, there are some drawbacks to the suggested framework. It makes the assumption of linear scalability, which might not be accurate in production environments that are dynamic or heterogeneous. Additionally, it disregards performance degradation brought on by model drift and does not account for qualitative output differences, treating all inferences as having equal value. Additionally, LCOAI currently ignores social externalities and environmental costs like energy use, carbon emissions, and workforce effects aspects that are becoming more and more important in the discussion of ethical AI and ESG compliance. Furthermore, even though the CAPEX and OPEX figures used in this study were based on publicly available pricing information, they might not accurately represent the cost variations that exist across regions, regulatory environments, and deployment sizes. These drawbacks imply that rather than being a universally applicable economic solution, LCOAI should be viewed as a foundational metric that encourages additional improvement and contextual adaptation.

Adopting standardized AI cost metrics, such as LCOAI, could be crucial from a policy and governance standpoint in enhancing accountability, transparency, and strategic alignment in AI investments made by the public and private sectors. Incorporating such lifecycle economic analyses into procurement frameworks, sustainability assessments, and guidelines for AI deployment may be advantageous for funding agencies and regulatory bodies. Furthermore, adding carbon cost per inference to LCOAI would help match AI investment choices with more general environmental and social objectives. LCOAI provides a significant step toward responsible, economical, and sustainable AI infrastructure planning balancing innovation with transparency, scale with stewardship as AI becomes more and more integrated into mission-critical systems.